\documentclass[english,twocolumn]{emulateapj}

\usepackage{epstopdf}

\newcommand\lsim{\mathrel{\rlap{\lower4pt\hbox{\hskip1pt$\sim$}}
    \raise1pt\hbox{$<$}}}
\newcommand\gsim{\mathrel{\rlap{\lower4pt\hbox{\hskip1pt$\sim$}}
    \raise1pt\hbox{$>$}}} 
\newcommand{\br}{\mathrm {b}}

\newcommand{\bc}{\mathrm{bc}}
\newcommand{\vbc}{\mathrm{vbc}}

\newcommand{\vrm}{{\rm v}}

\usepackage{amsmath}

\makeatother

\shorttitle{Stream velocity effects on structure formation}
\shortauthors{Naoz et al.}
\makeatother
\begin{document}



\title{Simulations of Early Baryonic Structure Formation with Stream Velocity: I.  Halo Abundance}

\author{Smadar Naoz\altaffilmark{1}, Naoki Yoshida\altaffilmark{2}, Nickolay Y. Gnedin\altaffilmark{3,4,5}  }

\altaffiltext{1}{ CIERA, Northwestern University, Evanston, IL 60208, USA}
\altaffiltext{2}{IPMU, University of Tokyo, 5-1-5 Kashiwanoha, Kashiwa, Chiba 277-8583, Japan}
\altaffiltext{3}{Particle Astrophysics Center, Fermi National Accelerator Laboratory, Batavia, IL 60510, USA}
\altaffiltext{4}{Kavli Institute for Cosmological Physics and Enrico Fermi Institute, The University of Chicago, Chicago, IL 60637 USA}
\altaffiltext{5}{Department of Astronomy \& Astrophysics, The University of Chicago, Chicago, IL 60637 USA}
\email{snaoz@northwestern.edu }
%

\begin{abstract}
It has been recently shown that the relative velocity between the dark
matter and the baryons ($v_\bc$) at the time of recombination can affect
the structure formation in the early universe \citep{Tes+10a}.  
We {\it statistically} quantify this effect using large cosmological 
simulations. 
 We use three
different high resolution sets of simulations (with separate transfer functions for baryons and
dark matter) that vary in box size,
particle number, and the value of the relative velocity between dark
matter and baryons. We show that the {\it total} number density of
halos is suppressed by $\sim 20\%$ at $z=25$ for $v_\bc=1\sigma_\vbc$,
where $\sigma_\vbc$ is the variance of the relative
velocity, while for $v_\bc=3.4\sigma_\vbc$ the relative suppression at
the same redshift reaches $50\%$, remaining at or above the $30\%$
level all the way to $z=11$.  We also find high abundance of 
``empty halos'', i.e., halos that have gas fraction below half of the cosmic
mean baryonic fraction $\bar{f}_\br$. Specifically we find that for
$v_\bc=1\sigma_\vbc$ {\it all} halos below $10^5$~M$_\odot$ are empty
at $z\geq 19$. The high abundance of empty halos
results in significant delay in the formation of gas rich mini-halos
and the first galaxies.
\end{abstract}

\section{Introduction}

Observations of the cosmic microwave background (CMB) show that the
Universe at cosmic recombination (redshift $z\sim 10^3$) was
remarkably uniform apart from small spatial fluctuations in the energy
density and in the gravitational potential
\citep[e.g.,][]{bennett}. The primordial inhomogeneities in the
density distribution grew over time and eventually led to formation of
galaxies, galaxy clusters and large-scale structure. Various
physical processes contributed to the growth of perturbations in dark
matter and baryons \citep[e.g.,][]{Peebles, pee, Ma}, which evolved
very differently before recombination. As the result, baryon density
fluctuations were much smaller than fluctuations in the dark matter
density at the time of recombination.
 
Recently \citet{Tes+10a} showed that not only the amplitudes of the
dark matter and baryon density fluctuations were different 
but also were their velocities. After
recombination, the sound speed of baryons dropped down to thermal
velocities, while the dark matter velocity remained unaffected - thus,
the relative velocity of the dark matter with respect to the baryons
become supersonic.  \citet{Tes+10a} also show that the relative
velocity between the dark matter and baryons is coherent on scales of
a few mega-parsec and is in the order of $\sim 30$~km~sec$^{-1}$ at the time of cosmic recombination. 
This relative velocity is often called the
``stream velocity'' in the literature, and throughout this paper we
will use this term. The stream velocity effect was previously
overlooked, because the velocity terms are actually of the second order
in perturbation theory. The stream
velocity effect leads to power suppression at mass scales that
correspond to masses of the first bound objects in the Universe
\citep[e.g.,][]{Yoshida+03early}. 
Using the Press-Schechter \citep{ps} formalism, 
\citet{Tes+10a} showed that the number density of halos is suppressed 
by more than $60\%$ for
halos with $M=10^6$~M$_\odot$ at $z=40$.  In a subsequent paper,
\citet{Tes+10b} also included the baryonic component following 
\citet{NB05}. They found that
the stream velocity also results in much higher minimum mass for a gas
rich halo at high redshifts, as compared to the case without the
stream velocity \citep[e.g.][]{NB07}.

\citet{Tes+10a} and \citet{Tes+10b} findings have prompted several  
studies on the stream velocity effect in numerical simulations.
\citet{Stacy+10} studied the collapse of gas in minihalos and found
only a mild delay ($\sim 10^7$~yr) for the time of collapse at low redshift. They 
also estimated the difference in redshifts of collapse for the minimum
halos that can collapse, with and without the stream velocity effects,
and found only minimal effect due to the thermal evolution of the gas collapsing into a minihalo at low redshifts. Their results also show that the delay of the structure formation becomes significant with higher redshift.   \citet{Maio+11} examined the influence
of the stream velocity on the formation of the smallest bound
objects. They showed that the abundance of mini-halos
($<10^7$~M$_\odot$) is reduced by a few percent and that the baryon
fraction in those halos is reduced by somewhat larger amount.
\citet{Dalal+10} showed that the characteristic imprint of the stream
velocity effect on the power spectrum of the earliest structures can
be used to distinguish the effects of minihalos on intergalactic gas
before and during reionization.  \citet{Greif+11} used high
resolution, moving-mesh simulations to study the effect of different
initial stream velocities on the virialisation of the gas in
minihalos; their simulations suggest that the stream velocity
substantially delay the onset of gravitational collapse in minihalos.
Although all these studies consistently show some delay in the
collapse of early gas clouds, a thorough statistical calculation is 
clearly needed to quantify these results, and, especially, to examine
critically the effect on early baryonic structure formation.
 
In this paper we quantify the suppression of structure formation due
to the stream velocity effect by means of cosmological simulations. We
use three different sets of high resolution simulations in order to
study the stream velocity effect \emph{systematically}, thus
understanding the overall picture (instead of concentrating on
specific halos). In a companion paper \citep[][in prep.]{Naoz+11b} we explore the
effect of stream velocity on the gas fraction in halos and compare
simulation results to the predictions from linear theory
\citep[e.g.][]{Tes+10b}. We use a set of simulations with different
box sizes, particle numbers, and the values for the stream velocity to
analyze the suppression of the structure formation as a function of
the stream velocity.

We first describe the parameters and initial conditions of our simulations 
(\S \ref{sim}).
We present our results and analysis of the halo mass functions in the 
simulations in Section \ref{sec:NofM}.
 Finally we offer a discussion in \S \ref{sec:dis}.

Throughout this paper, we adopt the following cosmological
parameters: ($\Omega_\Lambda$, $\Omega_{\rm M}$, $\Omega_b$, n,
$\sigma_8$, $H_0$)= (0.72, 0.28, 0.046, 1, 0.82, 70 km s$^{-1}$
Mpc$^{-1}$)  \citep{wmap5}.

\section{Cosmological Simulations}\label{sim}

\subsection{Simulation Setup}\label{sec:basic}

We use parallel $N$-body/hydrodynamics solver GADGET-2
\citep{Gadget,G2}. We describe the general features of the
simulations, which are also summarized in table \ref{table_sim}.
\begin{enumerate}
\item The first set, named ``$N=256$'', uses a total of $2\times256^3$
  dark matter and gas particles within a cubic box 
  of $200$~comoving kpc on a side. To realize statistically significant 
  number of halos
  in such a small box, we artificially increase gravitational
  clustering in the simulation by setting $\sigma_8=1.4$.  We choose
  this box size so that a $10^4$~M$_{\odot}$ halo is resolved with
  $\sim500$ particles - the value needed to estimate the halo gas
  fraction reliably \citep{NBM}.  The gravitational softening is set to be 
  $40$~comoving~pc,
  well below the virial radius of a $10^4$~M$_\odot$ halo
  ($\sim 680$~comoving~pc). All the simulations in this set are initialized at
  $z=199$.
\item The second set, named ``$N=512$'', uses a total of
  $2\times512^3$ dark matter and gas particles within a cubic box with
  the size of $700$~kpc. In this set we also artificially increase
  $\sigma_8$ to 1.4. With these parameters, a halo with 500 dark
  matter particles has a mass of $\sim 5 \times 10^4$~M$_\odot$. The
  softening length is set to be $68$~comoving~pc. All the simulations in this set
  are initialized at $z=199$.
\item The final set of simulations uses $2\times768^3$ dark matter and
  gas particles (which we name the ``$N=768$'' set) in a $2$~Mpc box,
  and starts at $z=99$. For these parameters a halo with $500$ dark
  matter particle has a mass of $\sim 10^5$~M$_\odot$. The softening
  length is set to be $0.2$ comoving kpc. We use the ``correct'' value of
  $\sigma_8=0.82$ for this simulation set.
\end{enumerate}
In  each simulation set, we explore a variety of the values for 
the stream velocity (see \S \ref{sec:vel} and table \ref{table_sim}). 


\begin{table}
 \caption{Parameters of the simulations}
\label{table_sim}
\begin{center}
\begin{tabular}{l c c c }
\hline
{SIM}&  $v_{\rm bc,0}$ & $\sigma_\vbc$ stream & Shift in      \\
  &   km~sec$^{-1}$&   velocity            & position  \\
\hline \hline
 & $256$ runs, & $0.2$~Mpc,& $z_{in}=199$\\
\hline \hline
256$_0$           & 0   & 0  & No   \\
256$_{1\sigma+p}$            & 5.8   &  1       &  Yes $+18.05$~kpc/h\\
256$_{1.7\sigma}$           & 10   &     1.7  & No   \\
256$_{1.7\sigma+p}$             & 10  &      1.7  & Yes  $+30.7$~kpc/h   \\
256$_{2.6\sigma}$            & 15   &  2.6       &  No \\
256$_{3.4\sigma}$            & 20  &  3.4      &  No \\
\hline\hline
& $512$ runs, &$0.7$~Mpc, & $z_{in}=199$\\
\hline\hline
512$_0$     &0     & 0  & No   \\
512$_{1\sigma}$     &   5.8   &1    & No \\
512$_{1.7\sigma}$     &   10 &1.7    & No \\
512$_{3.4\sigma}$     &   20 &3.4   & No \\
\hline\hline
& $768$ runs,&$2$~Mpc,& $z_{in}=99$\\
\hline\hline
768$_0$     &0     & 0  & No   \\
768$_{1\sigma}$     &   3   &1    & No \\
768$_{1.7\sigma}$     &   5  &1.7    & No \\
768$_{3.4\sigma}$         & 10     &     3.4  & No   \\
\hline
\vspace{-0.7cm}
\end{tabular}
\end{center}
\end{table}

\subsection{Initial Conditions}\label{sec:ICs}

As has been shown by \citet{NNB} and \citet{NB07}, setting up initial
conditions on the small spatial scales is a delicate issue. High
accuracy in initial conditions is crucial for accurately predicting
the halo mass function in the lowest mass regime ($M \la
10^7$~M$_{\odot}$).

Following \citet{Naoz+10}, we generate separate transfer functions for
dark matter and baryons as described in \citet{NB05}. Ideally, streaming
velocities should be realized in the initial conditions
in a self-consistent way with the transfer functions
that are calculated with streaming effect. However,
 these transfer functions do not take into account 
the stream velocity effect,  in consistency with the previous  calculations that included the stream velocity in simulations. For all runs,
glass-like initial conditions were generated using the Zel'dovich
approximation. For baryons, we have used a glass file with positions
shifted by a random vector, thus removing artificial coupling between nearby
dark matter and gas particles \citep{Yoshida03b}.  
We note that we have used the same
phases for dark matter and baryons in all of our simulations
except for those with position shift, see \S \ref{sec:posS}).


\subsection{Stream Velocity in the Simulations}\label{sec:vel}

\citet{Tes+10a} showed that, while the stream velocity varies in
space, its coherence length is quite large, many mega-parsecs. Hence, on scale
of our simulation boxes it can be treated as constant bulk motion of
baryons with respect to dark matter. We include the effect of stream
velocity by adding, at the initial redshift, an additional velocity to
the $x$ component of the baryons velocity vector. We test a range of
values for the stream velocity, which is convenient to quantify in
terms of its rms value on small scales, $\sigma_\vbc$. Specifically,
we test $v_\bc=1\sigma_\vbc$ through $v_\bc=3.4\sigma_\vbc$ for all
the simulations sets (see table \ref{table_sim}).

\subsection{Position Shift in the Initial Conditions}\label{sec:posS}

It is customary to set the initial conditions such that the phases of
dark matter and baryons are the same, as dictated by the linear
perturbation theory. However, because the stream velocity effect is of
the second order, it violates the assumption. Namely, since the
baryons move in the dark matter reference frame, the respective
density perturbations that were co-located at the time of
recombination, become separated in space. In other words, the phases
of dark matter and baryons at the same spatial location on
sufficiently small scales become unrelated.

The relative shift between dark matter and baryons is easily
calculated through the geodesic equation,
\begin{equation}
\label{eq:pec}
a \dot{x}=v_{\rm bc}(a) \ ,
\end{equation}
where $x$ is the comoving distance,  $a$ is the scale factor, and
$v_{\rm bc}(a)$ is the lagrangian velocity of a baryonic fluid element
in the dark matter reference frame as a function of the scale factor
$a$. Hence
\begin{equation}
\label{eq:x}
\Delta x_{\rm bc}(a)=\int^{t}_0\frac{v_{\rm bc}(a)}{a} dt \ ,
\end{equation}
where $t$ is the cosmic time at the scale factor $a$. Taking this
integral, we find the relative shift in the position between baryons
and dark matter at $z_{\rm in}=199$ is $\Delta x_{\rm bc} =
18.5$~comoving~kpc/h for $v_\bc=1\sigma_\vbc$ and is $\sim
30.7$~comoving~kpc/h for $v_\bc=1.7\sigma_\vbc$. We test the effect of
the position shift on two of the $N=256$ runs by adding the above
values to the $x$ component of the baryons at the initial redshift
$z_{\rm in}$, thus effectively changing the phases of the baryons
relative to dark matter.

As we find below, the relative shift between baryons and dark matter
is a small effect, and it does not invalidate the rest of our
simulations, where the position shift is neglected.

\subsection{Halo definition}

We locate dark matter halos by running a friends-of-friends
group-finder algorithm with a linking parameter of $0.2$ 
(only for the dark matter component). We use the
identified particle groups to find the center of mass of each halo. 
After the center is located, we calculate density profiles of dark matter and
baryons separately, assuming a spherical halo and using $2000$ radial
bins between $r_{\rm min}=0$~kpc and $r_{\rm max}=20$~kpc. Using the density profiles, we find the virial
radius $r_{vir}$ at which the total overdensity is $200$ times the
mean background density, and compute the mass and the gas fraction of each
halo within that radius.

Recently, \citet{More+11} showed that halos identified by the
friends-of-friends algorithm enclose an average overdensity that is
substantially larger than $200$ and its specific value depends on the
halo concentration. In our approach we use the friends-of-friends
algorithm only for finding the center of mass of a halo, and compute
the actual halo mass using the spherical overdensity of $200$.

We only retain halos that contain at least $500$ dark matter particles
within their virial radii. The choice allows us to estimate halo
masses to about 15\% precision \citep{Trenti_halos} and to estimate
halo gas fractions reliably to a similar level of accuracy
\citep{NBM}.

\section{Results}\label{sec:NofM}

\subsection{Suppression of the Halo Mass Function}

\begin{figure}
 \centering\plotone{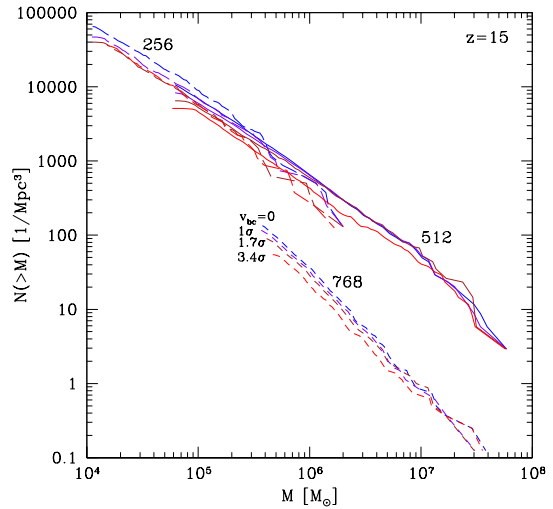}
\caption{Cumulative halo mass function for different simulations and valus
  of $v_{bc}$ at $z=15$. We show the $N=256$ (long dashed lines),
  $N=512$ (solid lines), and $N=768$ (short dashed lines). We consider
  the values for the stream velocity of $v_{bc}=0$,
  $v_\bc=1\sigma_\vbc$, $v_\bc=1.7\sigma_\vbc$, and
  $v_\bc=3.4\sigma_\vbc$, from top to bottom in each sets of lines.}
\label{fig:NofMz15}
\end{figure}

We first consider all the simulations that do not include the relative
position shift between baryons and dark matter (i.e., all runs except
$N=256_{1\sigma+p}$ and $N=256_{1.7\sigma+p}$).  \citet{Tes+10a}
calculated the number densities of collapsed halos with and without
the stream velocity, and illustrated the stream velocity effect on the
abundance of small halos. Specifically, based on the Press-Schechter
formalism, they showed that the number density of haloes with the rms
stream velocity $N_{\rm vbc} (>M)$ is suppressed by more than $60\%$
at the mass scale of $M\sim 10^6$~M$_\odot$ in comparison to the case
with no stream velocity ($N_0(>M)$). 

In Figure \ref{fig:NofMz15} we show the halo mass functions at $z=15$ for all
our simulations. Indeed, as can be seen, the halo number density is a
strong function of the stream velocity, resulting in substantial
suppression of the halo mass function for typical values of the stream
velocity.\footnote{The lower halo mass function for the $N=768$ runs
  is the result of our simulation setup; $N=768$ have much lower
  $\sigma_8$ than other simulation sets.}

Following \citet{Tes+10a}, we quantify the suppression of the halo
mass function as
\begin{equation}
  \label{eq:NofM}
  \Delta_{\rm v}=\frac{N_{\rm vbc} (>M)-N_{0} (>M)}{N_0(>M)}.
\end{equation}

A more relevant quantity is, perhaps, the abundance of halos that
contain substantial amount of gas. It is such halos that may host
first stars and first supernovae, and also serve as sinks of ionizing
radiation during reionization. Therefore, we also introduce
\begin{equation}
  \label{eq:NofMfg}
  \Delta_{{\rm v}, f_g}=\frac{N_{\rm vbc} (>M,f_g>\bar{f}_{\br}/2)-N_{0} (>M,f_g>\bar{f}_{\br}/2)}{N_0(>M,f_g>\bar{f}_{\br}/2)}
\end{equation}
as a relative difference between the mass functions of halos that
contain more gas than one half of the cosmic mean baryon fraction
($\bar{f}_{\br}$) in simulations with and without the stream velocity,
$N_{\rm vbc} (>M,f_g>\bar{f}_{\br}/2)$ and
$N_0(>M,f_g>\bar{f}_{\br}/2)$ respectively.

We would call halos with gas fractions less then $\bar{f}_{\br}/2$
``empty halos'', although, admittedly, this definition is somewhat
arbitrary and a different gas fraction threshold could also be
considered.

\begin{figure}
\centering\plotone{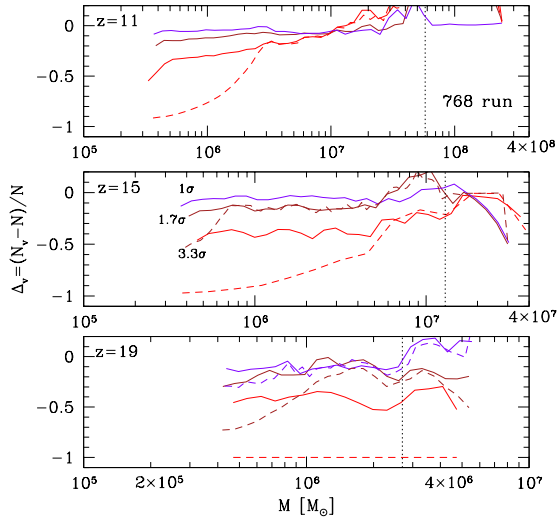}
\caption{Effect of the stream velocity on the cumulative halo mass
  function for the $N=768$ runs, for redshifts of (from bottom to top
  panels) 19, 15, and 11. The relative difference for all halos
  ($\Delta_\vrm$) is shown with the solid lines, while dashed lines
  show the relative difference for non-empty halos
  ($\Delta_{\vrm,f_g}$), for various values of the stream velocity
  $v_\bc=1\sigma_\vbc$, $v_\bc=1.7\sigma_\vbc$, and
  $v_\bc=3.4\sigma_\vbc$ (purple, brown and red lines respectively).
  Note that $\Delta_{\vrm,f_g}$ for $v_\bc=3.4\sigma_\vbc$ at $z=19$
  is exactly $-1$ (i.e., there are no halos with gas fraction above
  $\bar{f}_{\br}/2$ in that simulation). The opposite is true for
  $v_\bc=1\sigma_\vbc$ at $z=15$ and $z=11$ - there are no empty halos
  in those runs. The dotted vertical line in each panel marks the mass
  at which the halo number density is
  $5$~kpc$^{-3}$.} \label{fig:NEx768}
\end{figure}

\begin{figure}
\centering\plotone{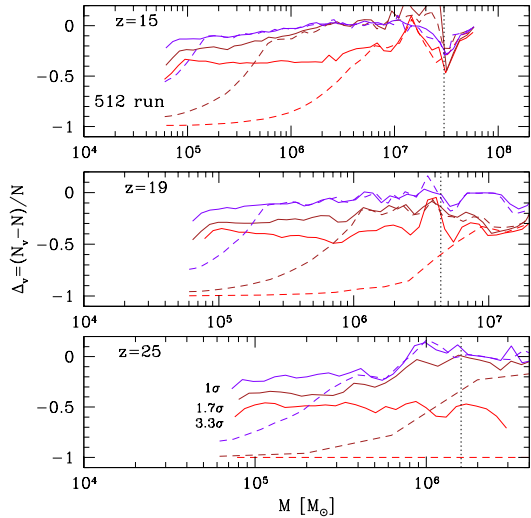}
\caption{Same as Figure \ref{fig:NEx768} for the $N=512$ runs at
  redshifts (from bottom to top panels) 25, 19, and 15.}
\label{fig:NEx512}
\end{figure}
  
First we show in Figure \ref{fig:NEx768} the relative suppression of
halo mass function due to the stream velocity for all and for
non-empty halos only ($\Delta_\vrm$ and $\Delta_{\vrm,f_g}$; solid and
dashed curves respectively) for the $N=768$ set of simulations. At
$z=19$ the suppression of the total number density of halos is about
$50\%$ for $10^6$~M$_\odot$ for $v_\bc=3.4\sigma_\vbc$ and $\sim 15\%$
for $v_\bc=1\sigma_\vbc$. As time goes by, the effect of the stream
velocity diminishes, but remains clearly visible even at the
lowest redshift we consider, $z=11$, still present.

Perhaps the most interesting feature shown in Figure
\ref{fig:NEx768} is the high abundance of empty halos. At $z=19$,
more than $20\%$ of the halos below $10^6$~M$_\odot$ are empty for
$v_\bc=1.7\sigma_\vbc$ and {\it all} of the halos in the simulation
are empty for $v_\bc=3.4\sigma_\vbc$. These empty halos lay above the
minimum cooling mass\footnote{We note that we in the simulations presented in the paper we included {\it only} adiabatic cooling. } \citep[$\sim 3\times10^5$~M$_\odot$,
  e.g.,][]{Trenti+09}. Thus, we conclude that, in patches of the
universe where the stream velocity is high, the formation of the first
stars and galaxies is delayed.  This behavior is consistent with
the $N=512$ simulation set (Figure \ref{fig:NEx512}), where, for
$v_\bc=3.4\sigma_\vbc$, most of the halos below $\sim 10^6$~M$_\odot$
are empty of baryons at $z=19$, and all of the halos in the simulation
are empty at $z=25$. Furthermore, for $v_\bc=1.7\sigma_\vbc$, more
than $50\%$ of the halos below $\sim 4\times10^5$~M$_\odot$ are empty
for $z\geq19$.

\begin{figure}
\centering\plotone{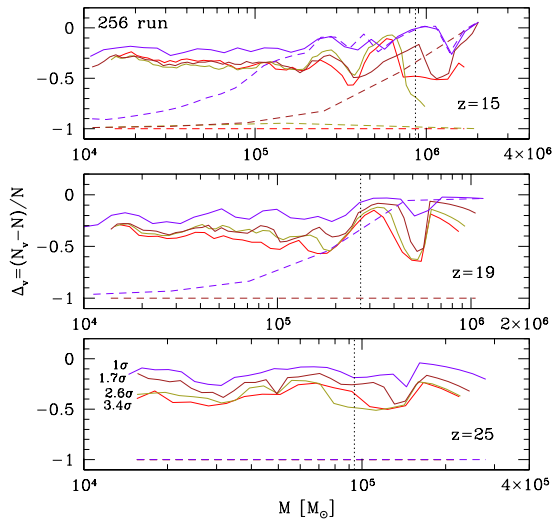}
\caption{Same as Figure \ref{fig:NEx768} for the $N=256$ runs at
  redshifts (from bottom to top panels) 25, 19, and 15 and for 4
  values of the stream velocity $v_\bc=1\sigma_\vbc$,
  $v_\bc=1.7\sigma_\vbc$, $v_\bc=2.6\sigma_\vbc$, and
  $v_\bc=3.4\sigma_\vbc$ (purple, brown, green and red lines
  respectively).}
\label{fig:NEx256}
\end{figure}

\subsection{Saturation of the Halo Mass Function Suppression in the Low
  Mass Limit}

\begin{figure}
\centering\plotone{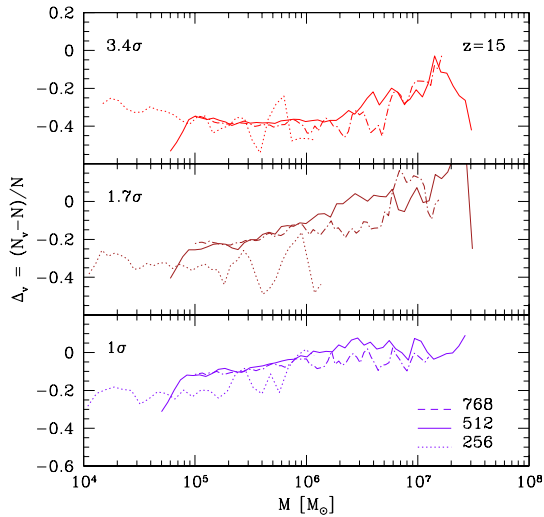}
\caption{Halo mass function suppression factor $\Delta_{\rm v}$ at
  $z=15$ for all our simulation sets for different values of the
  stream velocity (from bottom to top) of $v_\bc = 1\sigma_\vbc$,
  $1.7\sigma_\vbc$, and $3.4\sigma_\vbc$. In all panels $N=768$,
  $N=512$, and $N=256$ simulation sets are represented by dashed,
  solid, and dotted lines respectively. All 3 simulations sets are
  reasonably consistent with each other within the cosmic variance.
  Note that to check consistency with somewhat lower mass values, we
  have lowered the threshold for the number of particles per halo to $300$ particles per halos
  {\it only} for the $N=768$ runs and {\it
    only} in order to generate this figure. }
\label{fig:Nratioz15}
\end{figure}

As expected, larger values for the stream velocity result in stronger
suppression of the halo mass function. However, in the $N=256$
simulation set the effect clearly saturates for high values of
$v_\bc$ (see Figure \ref{fig:NEx256}). In order to explore the nature of this saturation further, we
compare $\Delta_{\rm v}$ at $z=15$ (the latest epoch that our $N=256$
runs can be continued to) between different simulation sets in Figure
\ref{fig:Nratioz15}.

Overall, the agreement between all three simulation sets is
reasonable\footnote{Note that this agreement is present through all the simulations despite the different $\sigma_8$ assumed.}; $N=256$ set suffers from poor statistics for $M \ga 10^5
{\rm M}_\odot$, so the agreement between that series and two other
sets is somewhat worse. Nevertheless, the effect of saturation for
low masses is clearly visible for all values of $v_\bc$.

The reason why $\Delta_{\rm v}$ as a function
of halo mass is expected to saturate at the low mass limit
is understood as follows. For halo
mass of $10^5$~M$_\odot$ the escape velocity is about
$0.77$~km~sec$^{-1}$, while the stream velocity for
$v_\bc=3.4\sigma_\vbc$ at $z=15$ is $1.6$~km~sec$^{-1}$ and at $z=25$
is $2.6$~km~sec$^{-1}$. Thus, it is not surprising that halos below
$10^5$~M$_\odot$ are empty in that redshift range - the stream
velocity is simply much larger then the halo escape velocity, so the
dark matter halo is unable to accrete any gas. Hence, such a halo will
have only about $(1-\bar{f}_{\br})$ of the mass it would have had if
$v_\bc=0$. Since the cumulative mass function $N(>M) \propto M^{-1}$
in this mass regime, the loss of baryons by low mass halos results in
$\bar{f}_{\br}\approx 0.17$ reduction in the halo mass function. 


At larger masses the interaction between the baryons moving with the
highly supersonic velocity and the dark matter halo becomes more
complex. For example, in the $v_\bc=3.4\sigma_\vbc$ case the mass
function is smaller by $\sim40\%$ at $10^5{\rm M}_\odot \la
M \la 3\times 10^6{\rm M}_\odot$. This magnitude of the effect cannot be
explained by the loss of baryons alone; some non-negligible fraction of
dark matter has to be lost by a halo as well. 
Note that the simulations sets $N=512$ and $N=768$ have 
different $\sigma_8$, but agree well  in the magnitude of the
halo mass function suppression. 
Hence, these results suggest that the suppression
effect is largely due to non-trivial non-linear 
interaction of the baryonic
flow and the dark matter halo.

%

%

\begin{figure}
\centering\plotone{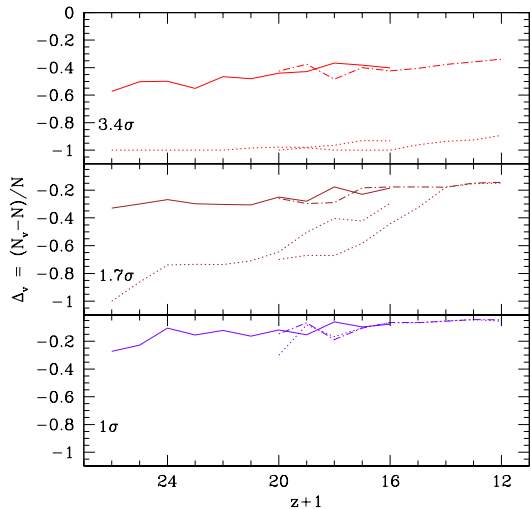}
\caption{Halo mass function suppression factor $\Delta_{\rm v}$ as a
  function of time for $5\times 10^5$~M$_\odot$ halos. In this figure
  we only show $N=768$ (dot-dashed curves) and $N=512$ (solid curves)
  simulation sets. We also show the suppression factor for non-empty
  halos, $\Delta_{\rm v,f_g}$, with dotted lines. We consider
  $v_\bc=1\sigma_\vbc$, $v_\bc=1.7\sigma_\vbc$, and
  $v_\bc=3.4\sigma_\vbc$ from bottom to top panels
  respectively.} 
\label{fig:Nratio_z}
\end{figure}

In Figure \ref{fig:Nratioz15} we consider the time evolution of the
halo mass function suppression factor for a halo mass that lays safely
above the molecular hydrogen cooling mass
\citep[e.g.,][]{Trenti+09}. We choose a value of $5\times
10^5$~M$_\odot$, which is well resolved and statistically reliably
represented in both $N=768$ and $N=512$ simulation sets.  This
comparison suggests significant delay in the formation epoch of gas rich
halos, which, in principle, could have formed stars. We find that for
$v_\bc=1\sigma_\vbc$ the suppression is modest, in agreement with
previous studies \citep{Maio+11,Stacy+10}. However, for larger values of
the stream velocity ($v_\bc=1.7\sigma_\vbc$ and
$v_\bc=3.4\sigma_\vbc$) we find a significant delay and large
abundance of empty halos.

\subsection{Position Shift}

\begin{figure}
\centering\plotone{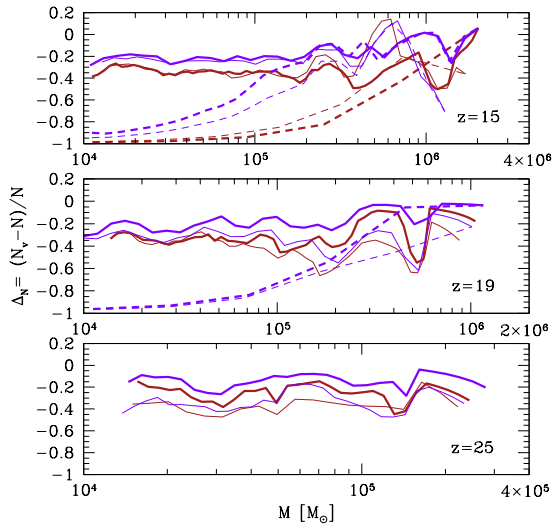}
\caption{Halo mass function suppression factors $\Delta_{\rm v}$ (solid
  curves) and $\Delta_{{\rm v},f_g}$ (dashed curves) for
  $v_\bc=1\sigma_\vbc$ and $v_\bc=1.7\sigma_\vbc$ cases (purple and
  brown curves respectively). Thin curves represent simulations that
  properly account for the position shift (256$_{1\sigma+p}$ and
  256$_{1.7\sigma+p}$ runs), while thick curves show our fiducial
  approach, without adding a position shift to the initial conditions
  (i.e., initializing the simulation with the same phases for the
  baryons and dark matter, 256$_{1\sigma_\vbc}$ and
  256$_{1.7\sigma}$). As one can see, the relative shift in the
  positions of baryons and dark matter makes only a modest
  correction.}
\label{fig:NofM}
\end{figure}

As we discussed in \S \ref{sec:posS}, the stream velocity also
causes the baryons to shift in position relative to dark mater.  We
have tested this effect in two of the $N=256$ simulation runs. In
Figure \ref{fig:NofM} we show the relative suppression of halo
formation comparing between the cases with and without the position
shift (thin and thick curves respectively). By $z=15$ the relative
shift in the positions is erased; at earlier times the
effect is noticeable, but it never becomes the dominant effect in
the change in the halo mass function induced by the stream velocity.

\subsection{Suppression of the  Clumping Factor}

The baryonic clumping factor plays an important role in  
the penetration and escape of radiation from an inhomogeneous 
medium and thus the effects of the stream velocity also 
depicted in the suppression of the clumping factor \citep[e.g.,][]{BL01}. 
Adopting \citet{SH03} for estimating the clumping 
factor $C$ from SPH simulations we define
\begin{equation}
C=\frac{\sum_i m_{b,i}\rho_i (<\Delta \rho_c)^{-1} \sum_j m_{b,j}\rho_j(<\Delta \rho_c) }{\left(\sum_k m_{b,k} \right)^{2}} \ ,
\end{equation}
where $m_b$ is the mass of the baryon particles  in our simulations,   
$\rho_c$ is the critical density in the Universe, 
and the summation is over all gas particles.

%

\begin{figure}
  \centering \includegraphics[width=84mm]{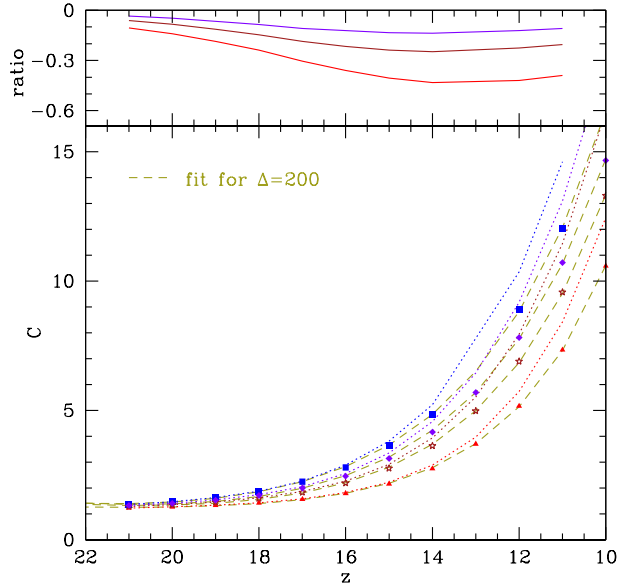}
\caption{The clumping factor as a function of redshift. In the bottom panel we consider the clumping factor evolution  for $v_\bc=0$, $v_\bc=1\sigma_\vbc$, $v_\bc=1.7\sigma_\vbc$, and
  $v_\bc=3.4\sigma_\vbc$ (blue,purple, brown and red lines respectively), choosing $\Delta=200$. We show in dashed lines the resulted fit to equation (\ref{eq:cfit}); see table \ref{table_fit} for the fit parameters.  In the top panel we consider the suppression ratio $C_\vbc/C_{\vbc=0}-1$ (see text)  for overdensity case of  $\Delta=200$ (we use the same color code as in previous figures). Note that the results for $z=13$ and $z=10$ of the run $768_0$ are missing due to computer failure.} \label{fig:clump}
\end{figure}

\begin{table}
 \caption{Fitting Parameters}
\label{table_fit}
\begin{center}
\begin{tabular}{l c c c c }
$v_\bc$ & $a_1$ & $a_2$ & $a_3$ & $a_4$ \\
\hline
\hline
$0$ & $415.1\pm66.4$ & $-0.32\pm0.01$ &  $0.06\pm0.12$ & $0.13\pm0.09$ \\
$1\sigma_\vbc$ & $ 393.8 \pm42.3$ & $-0.33\pm0.01$ &  $0.05\pm0.07$ & $0.15\pm0.08$ \\
$1.7\sigma_\vbc$ & $439.4\pm54.4$ & $ -0.35\pm0.01$ & $ 0.12\pm0.15$ & $0.1\pm0.06$ \\
$3.4\sigma_\vbc$ & $ 607.1 \pm75.7$ & $-0.41\pm0.01$ &  $0.55\pm0.27$ & $0.04\pm 0.02$ \\
\hline
\vspace{-0.7cm}
\end{tabular}
\end{center}
\end{table}

Following \citet{Maio+06} we sum over  all gas particles whose 
density $\rho_i$ is smaller than a given threshold $\Delta \rho_c$. 
Thus, we avoid numerical artifacts that are produced by gas particles 
belonging to collapsed objects which would artificially increase 
the value of $C$ due to their high density. 
In Figure \ref{fig:clump} we show the clumping factor for the $N=768$ run.  
We choose to show the results of only this set of  runs 
because the clumping factor depends on the assumed initial fluctuation
amplitudes.
For this run the gas particle mass is $m_b=225.5$~M$_\odot$ for all SPH 
particles\footnote{Unlike the variables $\Delta_v$ 
and $\Delta_{v,f_g}$ the clumping factor is very sensitive 
to the assumed initial fluctuation amplitudes 
(because essentially we count the number of clumps, 
which is larger for large $\sigma_8$).}.  
In Figure \ref{fig:clump}, 
bottom panel, we show the clumping factor for  
overdensity of $\Delta=200$, and we also test 
$\Delta=500$ (dotted lines).  Note, that we omit the $\Delta=100$ lines from the figure, to reduce the clutter, this choice gives a typical $1\%$ reduction of the clumping factor at $z=14$ and about $15\%$ at $z=10$ (for all $v_\bc$ cases). 
We also consider (top panel) the relative  
suppression ratio $C_\vbc/C_{\vbc=0}-1$, where $C_\vbc$ is 
the clumping factor for the case with stream velocity 
and $C_{\vbc=0}$ is for the case without the stream velocity effect. 

The growth of the clumping factor as a function of time 
is quite regular with a behavior, at high redshift, that resembles 
exponential asymptotic decay. 
We assume an exponential behavior and
find a simple fit in the form of: 
\begin{equation}
\label{eq:cfit}
C(z)=a_1 e^{a_2z}+a_3\exp^{a_4 z} \ .
\end{equation}
We summarize the fit results (for $\Delta=200$ for different values of $v_\bc$) in table \ref{table_fit}. 

As shown in Figure \ref{fig:clump} the stream velocity 
suppresses the clumping of baryon especially at lower redshift. 
At high redshift the clumping is similar and indeed low in all cases. 
However, as time goes by, the clumpiness in the case without 
stream velocity increases dramatically, 
while it is smaller by about $50\%$ for $v_\bc=3.4\sigma_\vbc$. 
Figure \ref{fig:clump} also shows that the suppression reaches 
a saturation at $z=14-12$ and below this redshift we find 
smaller differences. 

\section{Discussion}\label{sec:dis}

We have used three-dimensional hydrodynamical simulations to
investigate the effects of stream velocity on halo mass function in
the early universe \citep[in a companion paper ][we study the effect
  on the gas fraction and filtering mass]{Naoz+11b}. \citet{Tes+10a}
showed recently, within the frame work of linear theory, that the
initial velocity difference between baryons and dark matter after
recombination significantly suppresses the halo mass function at small
scales. 

We found that the {\it total} halo mass function as a function of mass is
significantly suppressed in all of our simulation sets (see Figures
\ref{fig:NEx768}--\ref{fig:NEx256}). Summarizing Figures
\ref{fig:NEx768}--\ref{fig:Nratio_z}, we find that for a range of
halos between $5\times 10^4-5\times 10^5$~M$_\odot$ the suppression of
the number density of halos at $z=25$ is $\sim 20\%$ for
$v_\bc=1\sigma_\vbc$, $\sim 40\%$ for $v_\bc=1.7\sigma_\vbc$, and
$\sim 55\%$ for $v_\bc=3.4\sigma_\vbc$. As time goes on, the
suppression of the halo mass function due to the stream velocity
diminishes with a varied rate - for example, by $z=15$ (the lowest
redshift we consider in this paper), for $v_\bc=1\sigma_\vbc$ the halo
mass function approaches the case with no stream velocity (i.e.,
no suppression), while in the $v_\bc=3.4\sigma_\vbc$ case the
suppression factor remains at $\sim 30\%$.

Our most significant result is the high fraction of halos that
are almost entirely
devoid of gas. We have found that for the extreme values of the stream
velocity we consider ($v_\bc=3.4\sigma_\vbc$), the gas simply does not
accrete on the dark matter halos - almost all halos below
$10^6$~M$_\odot$ throughout all of our simulation runs, are
empty. For more typical values of the stream velocity
($v_\bc=1\sigma_\vbc$) most of the empty halos are smaller than about
$10^5$~M$_\odot$ at $z\leq 19$.  Thus, we expect
significant delay in the formation of the first generation of
galaxies; in our companion paper \citet[][in prep.]{Naoz+11b} we study the mass
scales of halos that still retain their gas.


It was suggested in the literature that gas rich halos and minihalos
may play an important role in cosmic reionization, and that they can
produce distinct 21-cm signatures (\citet{Kuhlen,Shapiro+06,NB08} but see
\citet{Furlanetto06}). Minihalos can block ionizing radiation and
induce an overall delay in the initial progress of reionization
\citep[e.g.,][]{shapiro87,BL02, iliev+03,Shapiro+04,iss05,
  mcquinn07}. However, our results here \citep[and in our companion
  paper][in prep.]{Naoz+11b} suggest that at high redshift the stream velocity
effect results in a high abundance of empty halos halos below $\sim
10^5$~M$_\odot$.  Thus, if reionization started sufficiently early
\citep{Yoshida+07}, in patches of the universe where the stream 
velocity is
large, there are fewer gas rich halos that can absorb ionizing
photons.  Hence, in these patches the delay of the reionization
caused by minihalos may be less than in regions that happen to have a
small value of the stream velocity. The feature is clearly
quantified by using the gas clumping factor. 
The variable measure the clumpiness of baryonic structure {\it outside} 
of bound objects. We found that it is very sensitive to stream velocity 
effect (see Figure \ref{fig:clump}). 
Therefore, not only the formation of
the first generations of galaxies may be affected by the stream
velocity effect, but also the reionization process may proceed
differently in regions with very different stream velocities.

We note that recently \citet{Bittner+11}  showed
that if the ionizing sources were primarily in halos dominated
by atomic cooling (corresponding to $\gsim10^8$~M$_\odot$), then the effect on reionization is negligible.
However, for sources at high redshift, in halos dominated by molecular
hydrogen the effect on the evolution of the ionizing front
may be significant.


 
\section*{Acknowledgments}
We thank Avi Loeb, Rennan Barkana, Andrey Kravtsov, Neal Dalal, Will Farr and Dmitriy Tseliakhovich for
useful discussions.  We also thank Yoram Lithwick for the use of this
allocation time on the computer cluster Quest. This research was supported in part through the computational resources
and staff contributions provided by Information Technology at Northwestern
University as part of its shared cluster program, Quest. SN acknowledges support
from a Gruber Foundation Fellowship and from the National Post
Doctoral Award Program for Advancing Women in Science (Weizmann
Institute of Science). This work was supported in part by the DOE at
Fermilab, by the NSF grant AST-0908063, and by the NASA grant
NNX-09AJ54G. NY acknowledges financial support from 
from the Grant-in-Aid for Scientific Research (S) 20674003
by Japan Society for the Promotion of Science.
\bibliographystyle{hapj}
\bibliography{cosmo}

\end{document}